\documentclass[11pt]{article}
\usepackage{epsfig}
\usepackage{latexsym}
\usepackage{amsmath}
\usepackage{amsmath}
\usepackage{amsfonts,amssymb}
\newtheorem{theorem}{Theorem}[section]{\bf}{\it}
\newtheorem{lemma}[theorem]{Lemma}{\bf}{\it}
{\bf}{\it}
\newtheorem{corollary}[theorem]{Corollary}{\bf}{\it}

\renewcommand{\forall}{\mbox{for all}\,\,}
          \def\dt{\cal}
          \def\dA{{\dt A}}
          \def\dB{{\dt B}}

          \def\dM{{\dt M}}

          \def\C{{\cal C}}

          \def\H{{\cal H}}

          \def\K{{\cal K}}
          
          \def\M{{\cal M}}
          
          \def\O{{\cal O}}

          \def\T{{\cal T}}

          \def\W{{\cal W}}

          \def\gD{\Delta}

          \def\gO{\Omega}

          \def\eps{\varepsilon}

          \def\aloc{\dA_{\rm loc}}
          \def\aqloc{\widetilde{\dA}}
          \def\complex{{\bf C}}

          \def\Halmos{\quad\hfill$\Box$}
          
          \def\id{\rm id}
          \def\Im{{\rm Im}\,}

          \def\modop{\gD^{1/2}}
          
          \def\modopt{\gD^{it}}
          \def\modopmt{\gD^{-it}}
          \def\naturals{{\bf N}}
          \def\pct{P$_1$CT}

          \def\Rd{\reals^{1+s}}
          \def\reals{{\bf R}}

          \def\onu{\O_{\nu}}

\title{Two Uniqueness Results on the Unruh Effect and on PCT-symmetry}
\author{Bernd Kuckert\\
Mathematics Institute, Pl. Muidergracht 24\\
1018 TV Amsterdam, The Netherlands\\ e-mail:kuckert@science.uva.nl}
\date{September 2000}

\begin{document}
\maketitle

\begin{abstract}
The Unruh effect and a closely related form of PCT-symmetry have been
proved in general for finite-component Wightman fields by Bisognano
and Wichmann. While this result incorporates most of the fields
occurring in four-dimensional high energy physics, there still are 
field theories of interest
that are not covered (e.g., low-dimensional anyon fields and 
infinite-component fields). From the spectrum condition,
Borchers has derived a couple of commutation relations which
``almost, but only almost'' imply the Unruh effect and
PCT-symmetry. We show that this result does imply Unruh effect and
PCT-symmetry provided that the operators involved in Borchers'
commutation relations act geometrically on a local net of observables.
\end{abstract}

\section{Introduction}

Unruh's observation in the seventies
that the vacuum state of a free quantum field
appears as a temperature state
when looked at in a uniformly accelerated frame \cite{Unr76} has been proved
for all finite-component Wightman fields by Bisognano and Wichmann,
who addressed the problem around the same time and with a more mathematical
motivation \cite{BW75,BW76}. Starting from the field operators located in
the Rindler wedge, a generic algebraic construction yields
a strongly continuous one-parameter group of
unitary operators on the one hand and an antiunitary operator on the other.
Since these objects arise from the modular theory due to Tomita and Takesaki,
they are referred to as the {\em modular unitaries} and the {\em modular conjugation}
of the Rindler wedge,
respectively. The modular unitaries turned out to implement Lorentz
boosts, and the modular conjugations give rise to a PCT-operator.

On the other hand, the automorphism group implemented by the modular unitaries,
the {\em modular group}, exhibits a property that
characterizes thermodynamical equilibrium states, the so-called {\em KMS-condition}
(see, e.g., \cite{Haa92}), so the Bisognano-Wichmann result shows
that the vacuum state is a thermodynamical equilibrium
state not only in an inertial frame, but
also in a uniformly accelerated frame, as Unruh found independently
for the free field.

While the result of Bisognano and Wichmann is quite general, there
still are field theories of physical interest to which their arguments
do not apply, e.g., anyons in 1+2 dimensions or infinite-component
Wightman fields in any dimension.  What is more, there are examples of
theories that do not exhibit the Unruh effect
\cite{Yng94,BDFS98}, so the long term goal
is to find a criterion that tells the
theories showing the Unruh effect apart from those without this property.
An important application of the Bisognano-Wichmann symmetries in low dimensions,
which motivated the subsequent analysis, are a couple of recent proofs of
Pauli's spin-statistics connection and its extension to particles with
braid group statistics
\cite{GL94,GL95,Kuc95,Lon96,Mun98}. As this approach does not depend on
the special features of finite-component Wightman fields,
it also provides a spin-statistics connection for
1+2-dimensional theories without spinor structure. On the other hand,
the original Bisognano-Wichmann analysis is confined to finite-component Wightman
fields, whereas the symmetries they established are not confined to
these fields; so the further analysis of the Bisognano-Wichmann
symmetries is of interest on its own.

During the last decade,
several authors have been investigating the Unruh effect and PCT-symmetry in
the algebraic approach to relativistic quantum physics
\footnote{See Sect. \ref{conclusion} below for some examples,
see \cite{Ara99,Haa92} for textbooks on the algebraic approach.}.
This activity has been initiated by Borchers' proof of two
commutation relations which (essentially) imply the Unruh effect
for algebraic theories in 1+1 spacetime dimensions, and in higher
dimensions lead to the impression that there is not
much room left for theories violating the effect.

This impression has motivated our subsequent analysis.
The question is what can ``go wrong''  if
the modular unitaries or conjugation are known to implement a
``geometric action''  in some way to be made precise.

A first approach to this question specifies this ``geometric
action''  by assuming that the adjoint actions of the corresponding
operators map the algebra of observables
associated with a double cone region $\O\subset\Rd$
onto the algebra associated with some arbitrary open set $M_\O$. This
open set does, a priori, not need to be the image of the double cone
$\O$ under a function from $\Rd$ to $\Rd$. $M_\O$ may be bounded or
unbounded, and it may be connected or disconnected.  It has been shown
in \cite{Kuc97} that the modular conjugation implements a PCT-symmetry
provided that it implements any geometric action in this sense. If
the modular unitaries of the Rindler wedge act geometrically on the
net, they implement the boosts leaving the
wedge invariant {\em plus a possible translation along the edge of the
wedge}. It is shown below that this translation degree of freedom can
be eliminated by an application of the Borchers-Vladimirov double cone
theorem on analytic functions in several complex variables. That the
double cone theorem can be used here, was brought to my 
attention by S. Trebels, whose thesis contains similar results \cite{Tre97}. 
With this completion, the result provides a {\em first uniqueness theorem on
modular symmetries}.

A similar trick allowed the Unruh part of the problem to be treated in an
alternative way: if the modular unitaries of the Rindler wedge
map local observables onto local observables and let their localization
regions change in a continuous way, these regions have to transform as under
a boost which leaves the Rindler wedge invariant. This will be referred to as
the {\em second uniqueness theorem on modular symmetries}. In order to
investigate this question, the notion of a localization region of a single
local observable had to be made rigorous. This has been done in
\cite{Kuc98,Kuc99}.

This article is structured as follows: in Sect. \ref{results}, the
results are stated in precise terms and compared with each other.
The proofs follow in Sect. \ref{proofs}.  Some further
discussion also concerning related recent work is given in the
Conclusion.

\section{Preliminaries and Results}\label{results}

In what follows, $\dA$ denotes a local net of observables that
associates a unital C$^*$-algebra $\dA(\O)$ in an infinite-dimensional
Hilbert space $\H$ with each bounded open region $\O\subset\Rd$,
where $\Rd$ denotes a Minkowski spacetime with at least
two spatial dimensions. $\dA$ will be assumed to be {\bf isotonous},
i.e., $\dA(\O)\subset\dA(P)$ if $\O\subset P$, and to satisfy
{\bf locality}, i.e., if $\O$ and $P$ are spacelike separated open regions,
all elements of $\dA(\O)$ commute with all elements of
$\dA(P)$. For any unbounded open region $R$, one defines $\dA(R)$ to be the
C$^*$-algebra generated by the union of all $\dA(\O)$ for bounded open sets
$\O\subset R$. The elements of the union $\aloc$ of all $\dA(\O)$ associated
with bounded open regions $\O$ are called {\em local observables}.

Throughout this paper, the net $\dA$ will be assumed to exhibit the
following properties:
\begin{quote}
{\bf (A) Translation covariance.} There is a strongly continuous
unitary representation $U$ of the translation group $(\Rd,+)$ such that
for every bounded open region $\O\subset\Rd$, one has
$$U(a)\dA(\O)U(-a)=\dA(\O+a)\qquad\forall a\in\Rd.$$

{\bf (B) Spectrum condition.}
The (four-dimensional) spectrum of the four-momentum operator
generating the representation $U$ is a subset of
the closed forward light cone $\overline{V}_+$.

{\bf(C) Existence and uniqueness of the vacuum.}
The space of $U$-invariant vectors is one-dimensional. $\gO$ will
denote an arbitrary, but fixed unit vector in this space, the
{\em vacuum vector}. $\gO$ is {\em cyclic} with respect to the algebra
$\dA(\Rd)$.

{\bf (D) Reeh-Schlieder property.}
For every nonempty bounded open region $\O$, the vector
$\gO$ is cyclic with respect to the algebra $\dA(\O)$.
\end{quote}

Conditions (A) and (B) ensure that there is a well-defined four-momentum
operator whose energy spectrum is bounded from below, which makes the
system energetically stable. Conditions (A) through (C) are characteristic
for vacuum states in high energy physics.

Given Conditions (A) and (B), Condition (C) is equivalent to
irreducibility of the algebra $\dA(\Rd)$.
A necessary and sufficient condition for this is
that the bicommutant $\dA(\Rd)''$ of this algebra is a factor (Theorem
III.3.2.6 in \cite{Haa92}), and if $\H$ is separable, this can be assumed
without any loss of generality, since there always is a
direct-integral decomposition of $\aloc''$ into factors almost all
of which inherit the properties assumed so far (cf. also the remarks
in \cite{Haa92}, Sect. III.3.2, and references given there).

The Reeh-Schlieder property (Condition (D)) has
been established for all Wightman fields \cite{RS61},
and if, in the present setting, Conditions (A) through (C) hold,
Condition (D) is well known to hold if and only if one has
{\em weak additivity}, i.e., if for
each bounded region $\O$, one has $$\left(\bigcup_{a\in\Rd}\dA(\O+a)
\right)''=\aloc''.$$
For a proof that weak additivity is sufficient, see
\cite{Bor65}, or Thm. 7.3.37 in \cite{BaW92};
a simple proof that it is sufficient as well,
can be found, e.g., in \cite{TW97a,Kuc99}.

Since the vacuum vector is cyclic and, by locality, also separating
with respect to the von Neumann algebra $\dA(W_1)''$ of the {\em Rindler
wedge} $W_1:=\{x\in\Rd:\,x_1>x_0\}$, the map
$$A\gO\mapsto A^*\gO,\qquad A\in\dA(W_1)''$$
defines an antilinear operator
$S_{W_1}:\dA(W_1)''\gO\to\dA(W_1)''\gO$
which is closable. Its closure is called the
{\em Tomita operator} of $\gO$ and $\dA(W_1)''$ and admits a unique
polar decomposition $S_{W_1}=J_{W_1}\modop_{W_1}$ into an antiunitary
conjugation $J_{W_1}$ (the ``phase''  of $S_{W_1}$) which is
called the {\em modular conjugation} of $(\gO,\dA(W_1)'')$,
and a positive operator $\modop_{W_1}$ (the ``modulus'' 
of $S_{W_1}$) whose
square $\Delta_{W_1}$ is referred to as the {\em modular operator} of
$(\gO,\dA(W_1)'')$.

The main theorem of Tomita-Takesaki theory \cite{Tak70}
now implies that the adjoint actions of the
operators $\Delta_{W_1}^{it}$ map the algebras $\dA(W_1)''$ and
$\dA(W_1)'$ onto themselves, whereas the adjoint action of the
conjugation $J_{W_1}$ maps the two algebras onto one another.
Bisognano and Wichmann showed that for finite-component Wightman
fields, the unitary $\Delta_{W_1}^{it}$ coincides with the
unitary representing the
01-boost by $-2\pi t$ for all $t\in\reals$, whereas $J_{W_1}$
implements a charge conjugation together with a time reflection and a
spatial reflection in the 1-direction, this combination of discrete
transformations will be referred to as a {\em \pct-symmetry}.

For the algebraic setting, Borchers proved in \cite{Bor92}\footnote{For a
considerably simpler
proof found recently, see \cite{Flo98}.} that the spectrum
condition (without assuming Lorentz covariance) implies the
commutation relations
$$\begin{array}{lrcl}
(i) &J_{W_1}U(a)J_{W_1}&=&U(j_1a),\\
(ii) &\modopt_{W_1}U(a)\modopmt_{W_1} &=&U(\Lambda_1(-2\pi
t)a)\qquad\forall t\in\reals,
\end{array}$$
where $\Lambda_1(-2\pi t)$ denotes the Lorentz boost by $-2\pi t$
in the 1-direction, while $j_1$ is the reflection defined by
$$j_1x:=(-x_0,-x_1,x_2,\dots,x_s).$$
Wiesbrock noted that Borchers'
relations are not only a necessary, but also a sufficient condition
for the spectrum condition (\cite{Wie92}, cf. also \cite{BuS93}).
For 1+1 dimensions, Borchers' relations immediately imply
\cite{Bor92} that the net of observables may be enlarged to a local
net which generates the same wedge algebras (and hence the same
corresponding modular operator and conjugation) as the original one
and which has the property that $J_{_1}$ is a \pct-operator {\em
(modular \pct-symmetry)}, whereas $\modopt_{W_1}$ implements the
Lorentz boost by $-2\pi t$ for each $t\in\reals$ {\em (modular Lorentz
symmetry)}.

The first uniqueness theorem for modular symmetries states
that even in higher dimensions, $J_{W_1}$ or $\modopt_{W_1}$, $t\in\reals$,
can be shown to be a \pct-operator or a 0-1-Lorentz boost, respectively,
provided that $J_{W_1}$ or $\modopt_{W_1}$ implement {\em any} geometric action
on the net. The first step towards
it is the following lemma. In this lemma and in what follows, $\K$ will
denote the class of all double cones of the form
$\O:=(a+V_+)\cap(b-V_+)$, $a,b\in\Rd$.

\begin{lemma}\label{Hauptsatz}
Let $K$ be a unitary or antiunitary operator with the property that
for every double cone $\O$ there are open sets $M_\O$ and $N_\O$
such that
$$K\dA(\O)K^*=\dA(M_\O),\,\quad K^*\dA(\O)K=\dA(N_\O),$$
and let $\kappa$ be a causal
automorphism\footnote{Recall that a {\it causal automorphism}
of $\Rd$ is a bijection
$f:\,\Rd\to\Rd$ which preserves the causal structure of $\Rd$, i.e.,
$f(x)$ and $f(y)$ are timelike with respect to each other
if and only if $x$ and $y$ are timelike with respect to each other.
Without assuming linearity or continuity, one
can show that the group of all causal automorphisms of $\Rd$ is
generated by the elements of the Poincar\'e group and the dilatations
\cite{Ale50,AlO53,Ale75,Zee64,BoH72}.
Since the transformations implemented on the
translations by Borchers' commutation relations happen to be causal
in all applications discussed below,
this assumption means no loss of generality.} of $\Rd$ such that
$$KU(a)K^*=U(\kappa a)\qquad\forall a\in\Rd.$$
Then there is a unique $\xi\in\Rd$ such that
$$K\dA(\O)K^*=\dA(\kappa\O+\xi), \qquad \forall \O\in\K.$$
\end{lemma}
A first proof of Lemma \ref{Hauptsatz} was published in \cite{Kuc97}, but
both the statement and the proof given there were more
general, which made the formulation somewhat
technical. For the reader's convenience a less general, but more
accessible formulation is used here, and a more detailed version of
the proof is given below.

The following theorem is a consequence of
Lemma \ref{Hauptsatz} and Borchers' commutation relations.

\begin{theorem}[First Uniqueness Theorem]\label{mod sym}

(i) If for every double cone $\O\in\K$ there is an open set $M_\O$
such that
$$J_{W_1}\dA(\O)J_{W_1}=\dA(M_\O),$$
then
$$J_{W_1}\dA(\O)J_{W_1}=\dA(j_1\O)\qquad\forall \O\in\K.$$

(ii) If for every $t\in\reals$ and for
every $\O\in\K$ there is an open set $M^t_\O$ such that
$$\modopt_{W_1}\dA(\O)\modopmt_{W_1}
=\dA(M^t_\O),$$
then
$$\modopt_{W_1}\dA(\O)\modopmt_{W_1}=\dA(\Lambda_1(-2\pi t)\O)\qquad\forall\O\in\K.$$
\end{theorem}
\bigskip\noindent%
The statement of part (ii) implies the statement of part (i) \cite{GL94},
i.e., the Unruh effect implies modular \pct-symmetry. Further results
relating the above statements to each other and to similar conditions
can be found in \cite{Dav95}.

Assuming that $\gO$ is separating with respect to the algebra $\dA(V_+)$,
Borchers also found commutation relations for
the corresponding modular conjugation and unitaries: for each $a\in\Rd$,
he found that
\begin{align*}
J_+U(a)J_+&=U(-a);\\
\modopt_+U(a)\modopmt_+&=U(e^{-2\pi t}a)\qquad\forall t\in\reals.
\end{align*}
These relations, together with
Lemma \ref{Hauptsatz}, imply the following corollary:

\begin{corollary}[Uniqueness Theorem ``1a'']\label{mod sym+}

Assume $\dA$ to be Poincar\'e covariant, and assume
that the vacuum vector $\gO$ is
separating with respect to the algebra $\dA(V_+)''$, and
let $\modopt_{V_+}$ and $J_{V_+}$ be the corresponding modular
operator and conjugation, respectively.

(i) If for every double cone $\O$ there is an open set $M_\O$ such that
$$J_{V_+}\dA(\O)J_{V_+}=\dA(M_\O),$$
then
$$J_{V_+}\dA(\O)J_{V_+}=\dA(-\O)\qquad\forall \O\in\K.$$

(ii) If for every $t\in\reals$ and every double cone $\O$ there is an
open set $M_\O^t$ such that
$$\modopt_{V_+}\dA(\O)\modopmt_{V_+}=\dA(M_\O^t),$$
then
$$\modopt_{V_+}\dA(\O)\modopmt_{V_+}=\dA(e^{-2\pi t}\O)\qquad\forall\O\in\K.$$

\end{corollary}
\bigskip\noindent%
Since massive theories cannot be dilation invariant unless their mass
spectrum is dilation invariant (cf., e.g., \cite{MS69}),
the models concerned by part (ii) of this corollary
are massless theories. But it follows from the
scattering theory for massless fermions and bosons in 1+3
or 1+1 dimensions (see \cite{Buc75,Bu75a,Buc77}) that
either of the symmetry properties found in part (i) and part (ii)
of the corollary
implies a massless theory to be free (i.e., its S-matrix is trivial)
(see \cite{Bu75a,Bu77a,BF77}).

Note that for the 1+1-dimensional case,
all modular symmetries considered in Thm.
\ref{mod sym} and Cor. \ref{mod sym+} have been
established in \cite{Bor92}.

It is assumed above that the adjoint actions of
$J_{W_1}$ and $\modopt_{W_1}$, $t\in\reals$, map each local
algebra $\dA(\O)$, $\O\in\K$, {\em onto} the algebra $\dA(M_\O)$
associated with some open region $M_{\O}$
in Minkowski space. This means that, essentially, the net
structure has to be preserved. This is the restrictive aspect of the
assumption. On the other hand, the shape of the region $M_{\O}$
is left completely arbitrary, the map
$\K\ni\O\mapsto M_{\O}$ is not even assumed to be
induced by a point transformation. In this aspect, the
above assumptions are rather weak.

But there are, of course, other ways to specify what a ``geometric
action''  is. Denote by
$\W$ the class of all {\em wedges},
i.e., all images of the Rindler wedge $W_1$ under Poincar\'e
transformations. For $M\subset\Rd$, define the causal complement
$M^c$ to be the set of all points that are spacelike to $M$, and let
$M'$ denote the interior of $M^c$.
It has been shown in \cite{Kuc98,Kuc99} that one can define
a nonempty localization region for each local
observable $A\notin\complex\id$ by
$$L(A):=\bigcap\{\overline{W}:\,W\in\W,\,A\in\dA(W')'\}.$$
This localization prescription will be said to satisfy {\em locality}
if any two local observables $A$ and $B$ with the property that
$L(A)$ and $L(B)$ are spacelike separated commute. This property
does not follow from the locality property of the net alone, but with the
following additional assumptions one can derive it for the present setting
\cite{Kuc99}:
\begin{quote}
{\bf (E) Wedge duality.} $\dA(W')'=\dA(W)''$ for each wedge $W\in\W$.

{\bf (F) Wedge additivity.}  For each wedge $W\in\W$ and each double
cone $\O\in\K$ with $W\subset W+\O$ one has
$$\dA(W)''\subset\left(\bigcup_{a\in W}\dA(a+\O)\right)''.$$
\end{quote}
Wedge duality is a property of all finite-component Wightman fields by
the Bisognano-Wichmann theorem, and wedge additivity is a standard
property of Wightman fields as well. Condition (F) is
slightly stronger than the definition of wedge additivity used
in \cite{TW97a,Kuc99}, where the algebras $\dA(a+\O)$ in Condition (F)
are replaced by the larger algebras $\dA(a+\O')'$, but as this
difference is not expected to be substantial for physics, we use the
same term for convenience, which is in harmony with the other existing
notions of additivity used in algebraic quantum field theory.

Assume now that the localization region of the observable
$A_t:=\modopt_{W_1}A\modopmt_{W_1}$ depends continuously on $t$, i.e.,
that for every sequence $(t_\nu)_{\nu\in\naturals}$ which converges to
some $t_\infty\in\reals$, the localization region $L(A_{t_\infty})$
consists precisely of all accumulation points of sequences
$(x_\nu)_{\nu\in\naturals}$ with $x_\nu\in L(A_{t_\nu})$.

Then the following lemma establishes a first restriction on how the
localization region can depend on $t$.

\begin{lemma}\label{intermezzo}

With Assumptions (A) -- (E), suppose the localization prescription
$L$ defined above satisfies locality. Let $A$ be a local
observable in $\dA(W_1)$, and assume that there exists an
$\eps>0$ such that all $A_t$, $t\in[0,\eps]$, are local
observables and such that the function $[0,\eps]\ni t\mapsto
L(A_t)$ is continuous in the above sense. Then
\begin{quote}
(i) $L(A_{\eps})\subset \Lambda_1(-2\pi \eps)
\left(\overline{(L(A)+W_1)^{cc}}\cap\overline{(L(A)-W_1)^{cc}}\right)$;

(ii) $L(A_\eps) \subset L(A)-\overline{V}_+$;

(iii) $L(A)\subset L(A_\eps)+\overline{V}_+$.

\end{quote}
\end{lemma}
It is shown in the Appendix that the continuity assumption made 
on $t\mapsto L(A_t)$ is equivalent to continuity with respect to a metric 
first considered by Hausdorff, and that $\bigcup_{t\in[0,\eps]}L(A_t)$
is compact.

Next suppose that $t\mapsto L(A_t)$ is continuous not only for
sufficiently small $t$, but for all $t\in\reals$, and assume
wedge additivity in addition. With these slightly strengthened
assumptions one can now prove the following:
\begin{theorem}[Second Uniqueness Theorem]\label{second}

With Assumptions (A) -- (F), assume that
$\modopt_{W_1}\aloc\modopt_{W_1}=\aloc$, and suppose that
$L(A_t)$ depends continuously from $t$ for all $t\in\reals$ and
for all $A\in\aloc$. Then
$$L(\modopt_{W_1}A\modopmt_{W_1})
=\Lambda_1(-2\pi t)L(A)\qquad\forall A\in\aloc.$$

\end{theorem}
By the result of Guido and Longo, the conclusion of this proposition
also implies modular \pct-symmetry, but Proposition \ref{second} does not
provide a proper parallel to the \pct-part of the first uniqueness
theorem, which may also apply if the modular group does not act
in any geometric way.

The assumption that every local observable $A$ is mapped
onto some other local observable under the adjoint action of the
modular group prevents $A$ to
be mapped onto an observable localized in an unbounded
region. For every bounded open region $\O$ there are conformal
transformations which map $\O$ onto an unbounded region; these
transformations are excluded a priori. In contrast, the assumptions
of the first uniqueness theorem do not exclude these symmetries
explicitly, while it is evident from this theorem
that the modular objects under consideration cannot implement
these symmetries.

Another restrictive assumption of the second uniqueness
theorem is that wedge duality is assumed there,
whereas the first one can be used to derive wedge duality.
On the other hand the assumptions made in the second uniqueness theorem
admit the situation that
the net structure of $\dA$ is destroyed completely under the action of
the modular group.

\section{Proofs}\label{proofs}
For every algebra $\dM\subset\dB(\H)$, define its {\bf localization
region $L(\dM)$ with respect to the net $\dA$} by
$$L(\dM):=\{\O\in\K:\,\dA(\O)\subset\dM\}.$$
The only reason to use the class $\K$ of double cones
in this definition is convenience; one could replace $\K$ by the
larger class $\T$ of all open sets in $\Rd$ without affecting the
definition. To see this, denote the localization region obtained this
way by $L_\T(\dM)$; it is trivial that $L(\dM)\subset L_\T(\dM)$ as
$\K\subset\T$, while from isotony of
the net and the fact that each open region $M$ is the union of all
double cones $\O\subset M$, one finds
\begin{align*}
L_\T(\dM)&=\bigcup\{M\in\T:\,\dA(M)\subset\dM\}\\
&=\bigcup\{\O\in\K:\,\exists M\in\T:\,\O\subset\M,\dA(M)\subset\dM\}\\
&\subset\bigcup\{\O\in\K:\,\dA(\O)\subset\dM\}=L(\dM),
\end{align*}
which is the converse inclusion.

It is obvious from the definitions that $L(\dA(M))\supset M$.
For causally complete and convex regions one can prove the converse
inclusion, which we recall without proof from \cite{Kuc99} (Cor. 5.4)
for later use. Here a {\em causally complete} region is a region $R$
such that $(R^c)^c=R$.

\begin{lemma}\label{Landau2}

Let $R\subset\Rd$ be a causally complete convex open region.

(i) For every open region $M\subset\Rd$, one has $\dA(M)\subset\dA(R')'$
if and only if $M\subset R$.

(ii) $L(\dA(R))=R$.

\end{lemma}
One also checks that for any such $R$, one has
$L(\dA(R))=L(\dA(R)'')=L(\dA(R')')$.
We emphasize that the above assumption $s\geq 2$ is crucial for this
lemma; in 1+1 dimensions, there are chiral theories which do not obey
the statement of the lemma. The repeated use of this lemma in the proofs is
the main reason why $s\geq2$ is assumed throughout this paper.

\subsection*{Proof of Lemma \ref{Hauptsatz}}\label{proof Hauptsatz}

In what follows, $K$ and $\kappa$ are defined as in Lemma
\ref{Hauptsatz}. As before, $\K$ will denote the class of double cones.
For any open region $M\subset\Rd$, we denote by $\K^M$ the class of
all double cones $\O\in\K$ with $\O\subset M$, and for each
subalgebra $\dM$ of $\dB(\H)$, we denote by $\K^\dM$ the class
of all double cones $\O$ such that $\dA(\O)\subset\dM$.

The proof will be subdivided into five lemmas. The first
implies that for every $\O\in\K$, the regions $M_\O$ and $N_\O$ are
bounded. It uses the fact that a region $M$ is bounded if and
only if its difference region $M-M$ is bounded, and that
difference sets can be expressed in terms of translations. Since the
behaviour of translations under the action of the symmetry $K$ is
known by assumption, one can prove the following lemma.

\begin{lemma}\label{difference sets}

For every double cone $\O\in\K$, one has
$$L(K\dA(\O)K^*)-L(K\dA(\O)K^*)=\kappa(\O-\O).$$

\end{lemma}
{\it Proof.}
Using the assumptions of Theorem \ref{Hauptsatz}, one obtains
\begin{align*}
L&(K\dA(\O)K^*)-L(K\dA(\O)K^*)=L(\dA(M_\O))-L(\dA(M_\O))\\
&=\{a\in\Rd:\,\exists
P\in\K^{\dA(M_\O)}:\,\dA(P+a)\subset\dA(M_\O)\}\\
&=\{a\in\Rd:\,\exists P\in\K^{\dA(M_\O)}:\,
KU(\kappa^{-1}a)K^*\dA(P)KU(-\kappa^{-1}a)K^*\subset\dA(M_\O)\}\\
&=\kappa\{a\in\Rd:\,\exists P\in\K^{\dA(M_\O)}:\,
U(a)\underbrace{K^*\dA(P)K}_{=\dA(N_P)}U(a)
\subset\underbrace{K^*\dA(M_\O)K}_{=\dA(\O)}\}\\
&\subset\kappa\{a\in\Rd:\,\exists P\in\K^{\dA(M_\O)}:\,
\exists Q\in\K^{\dA(N_P)}:\,
\dA(Q+a)\subset\dA(\O)\}.
\end{align*}
Since the definitions and isotony imply
$$\K^{\dA(N_P)}=\K^{K^*\dA(P)K}\subset\K^{K^*\dA(M_\O)K}
=\K^{\dA(\O)},$$
and since, as remarked above, $\K^{\dA(\O)}=\K^\O$, one obtains
\begin{align*}
L(K\dA(\O)K^*)-L(K\dA(\O)K^*)&\subset\kappa\{a\in\Rd:\,\exists
Q\in\K^\O:\,\dA(Q+a)\subset\dA(\O)\}\\&
=\kappa(\O-\O).
\end{align*}
Conversely,
\begin{align*}
\kappa(\O-\O&)
=\kappa\{a\in\Rd:\,\exists
P\in\K^\O:\,\dA(P+a)\subset\dA(\O)\}\\&
=\{a\in\Rd:\,\exists P\in\K^\O:\,\dA(P+\kappa^{-1}a)\subset\dA(\O)\}\\&
=\{a\in\Rd:\,\exists P\in\K^\O:\,
K^*U(a)K\dA(P)K^*U(-a)K\subset\dA(\O)\}\\&
=\{a\in\Rd:\,\exists P\in\K^\O:\,
\dA(M_P+a)\subset\dA(M_\O)\}\\&
\subset\{a\in\Rd:\,\exists P\in\K^\O:\,\exists Q\in\K^{\dA(M_P)}:\,
\dA(Q+a)\subset\dA(M_\O)\},
\end{align*}
and since
$$\K^{\dA(M_P)}=\K^{K\dA(P)K^*}\subset\K^{K\dA(\O)K^*}=\K^{\dA(M_\O)},$$
one obtains
\begin{align*}
\kappa(\O-\O)&\subset\{a\in\Rd:\,\exists Q\in\K^{\dA(M_\O)}:\,
\dA(Q+a)\subset\dA(M_\O)\}\\
&=L(\dA(M_\O))-L(\dA(M_\O)).
\end{align*}

\Halmos

\bigskip\bigskip\noindent%
The next lemma proves that strict inclusions of double cones
are preserved under the adjoint action of the operator
$K$. Again, this boils down to translating local algebras
up and down Minkowski space and using the commutation relations
between $K$ and the translation operators.
One uses the fact that
$\overline{\O}\subset P$ if and only if $\O$ can be translated within $P$
into all directions.

\begin{lemma}\label{inclusions}

For any two double cones $\O,P\in\K$
with $\overline{\O}\subset P$, one has
$$\overline{L(K\dA(\O)K^*)}\subset L(K\dA(P)K^*).$$

\end{lemma}
{\it Proof.} $\overline{\O}\subset P$ if and only if the set
$\{a\in\Rd:\,\O+a\subset P\}$
is a neighbourhood of the origin of $\Rd$. After using Lemma
\ref{Landau2}, elementary transformations yield
\begin{align*}
\{a\in\Rd:\,\O+a\subset&P\}=\{a\in\Rd:\,\dA(\O+a)
\subset\dA(P)\}\\
&=\{a\in\Rd:\,K^*U(\kappa a)K\dA(\O)K^*U(-\kappa a)K
\subset \dA(P)\}\\
&=\{a\in\Rd:\,\dA(M_\O+\kappa a)\subset \dA(M_P)\}\\
&=\kappa^{-1}\{a\in\Rd:\,\dA(M_\O+a)\subset\dA(M_P)\}\\
&\subset\kappa^{-1}\{a\in\Rd:\,L(\dA(M_\O))+a\subset L(\dA(M_P))\}.
\end{align*}
Since $\kappa$ is a linear automorphism of $\Rd$, it
follows that $\overline\O$ can be a subset of $P$ only if
$$\{a\in\Rd:\,L(\dA(M_\O))+a\subset L(\dA(M_P))\}$$
is a neighbourhood of the origin. This implies the statement.
\Halmos

\bigskip\bigskip\noindent%
The next lemma proves that the maps
$$\K\ni\K\mapsto L(K\dA(\O)K^*)$$
and
$$\K\ni\O\mapsto L(K^*\dA(\O)K)$$
are induced by continuous functions $\tilde\kappa:\Rd\to\Rd$
and $\hat\kappa:\Rd\to\Rd$.
\begin{lemma}\label{bases}

Let $x\in\Rd$ be arbitrary, and
let $(\onu)_{\nu\in\naturals}$ be a neighbourhood base of
$x$ consisting of double cones $\onu\in\K$.

Then $(L(K\dA(\onu)K^*))_{\nu\in\naturals}$
is a neighbourhood base of a
(naturally, unique) point $\tilde\kappa(x)\in\Rd$, and
$(L(K^*\dA(\onu)K))_{\nu\in\naturals}$ is a neighbourhood base of
a point $\hat\kappa(x)\in\Rd$. The functions
$x\mapsto\tilde\kappa(x)$
and $x\mapsto\hat\kappa(x)$
are continuous.

\end{lemma}
{\it Proof.}
Without loss of generality, one may assume that
$\overline{\O}_{\nu+1}\subset\onu$ for all
$\nu\in\naturals$. It follows from
$L(\dA(\O))=\O\quad\forall\O\in\K$ and Lemma \ref{difference sets}
that all $L(K\dA(\onu)K^*)$, $\nu\in\naturals$, are bounded sets,
and it follows from Lemma \ref{inclusions} that
$$\overline{L(K\dA(\O_{\nu+1})K^*)}\subset L(K\dA(\onu)K^*).$$
Therefore, the intersection of this family is nonempty,
and Lemma \ref{difference sets} implies that the
diameter of $L(K\dA(\onu)K^*)$ tends to zero as $\nu$ tends to infinity.
This implies that the intersection contains precisely one point
$\tilde\kappa(x)$,
as stated. The corresponding statements for $K^*$ are proved
analogously.

This proves that $x\mapsto\tilde\kappa(x)$ is a bijective
point transformation. Let $(x_\nu)_{\nu\in\naturals}$
be a sequence in $\Rd$ that converges to a
point $x_\infty$. Then there is a neighbourhood base
$(\onu)_{\nu\in\naturals}$ of $x_\infty$ with $x_\nu\in\onu$ for all
$\nu\in\naturals$. But since $\tilde\kappa(x_\nu)\in
\tilde\kappa(\onu)$ for all $\nu\in\naturals$, and since
$\tilde\kappa(\onu)$ is a neighbourhood base of
$\tilde\kappa(x_\infty)$, it follows that $\tilde\kappa(x_\nu)$ tends
to $\tilde\kappa(x_\infty)$ as $\nu\to\infty$. This line of argument
applies to $\hat\kappa$ as well.
\Halmos

\bigskip\bigskip\noindent%
The next lemma determines the functions $\tilde\kappa$ and $\hat\kappa$
up to a constant translation.

\begin{lemma}\label{causal}

For every $x\in\Rd$, one has
$$\tilde\kappa(x)=\tilde\kappa(0)+\kappa x,$$
and
$$\hat\kappa(x)=\hat\kappa(0)+\kappa^{-1} x.$$

\end{lemma}
{\it Proof.} Let $(\onu)_{\nu\in\naturals}$ be a neighbourhood
base of $o$. Then $(\onu+x)_{\nu\in\naturals}$ is a neighbourhood
base of $x$, and
$$\bigcap_{\nu\in\naturals}L(K\dA(\onu+x)K^*)=\bigcap_{\nu\in\naturals}
\tilde\kappa(\onu+x)=\{\tilde\kappa(x)\}.$$
On the other hand,
\begin{eqnarray*}
\bigcap_{\nu\in\naturals}L(K\dA(\onu+x)K^*)&=&\bigcap_{\nu\in\naturals}
L(U(\kappa x)K\dA(\onu)K^*U(-\kappa x))\\
&=&\kappa x+\bigcap_{\nu\in\naturals}\tilde\kappa(\onu)\\
&=&\kappa x+\{\tilde\kappa(0)\}.
\end{eqnarray*}
The corresponding reasoning also leads to the statement made on $\hat\kappa$.

\Halmos

\bigskip\bigskip\noindent%
It has been shown now that $L(K\dA(\O)K^*)=\tilde\kappa(\O)$
for each double cone $\O\in\K$, and since $K\dA(\O)K^*=\dA(M_\O)$
by assumption, one concludes from $M_\O\subset K(\dA(M_\O))$ and isotony
that
$$K\dA(\O)K^*\subset\dA(\tilde\kappa(\O))\qquad\forall\O\in\K$$
and that
$$K^*\dA(\O)K\subset\dA(\hat\kappa(\O))\qquad\forall\O\in\K.$$
Using this, one can now prove that $\tilde\kappa$ and $\hat\kappa$ are inverse to
each other.
\begin{lemma} $\hat\kappa=\tilde\kappa^{-1}$,
and in particular, $\tilde\kappa$ and $\hat\kappa$ are
homeomorphisms.

\end{lemma}
{\it Proof.} For every double cone $\O$,
it follows from the preceding results that
$$\dA(\O)=K^*K\dA(\O)K^*K\subset K^*\dA(\tilde\kappa(\O))K\subset
\dA(\hat\kappa(\tilde\kappa(\O))),$$
and since $\hat\kappa(\tilde\kappa(\O))$ is a double cone by Lemma \ref{causal},
one can use Lemma \ref{Landau2} to conclude that $\O\subset\hat\kappa(
\tilde\kappa(\O))$. On the other hand, it follows from Lemma \ref{difference sets}
that the radii of the double cones $\O$ and $\hat\kappa(\tilde\kappa(\O))$ are
equal, so these double cones coincide, and as this applies for any double cone
$\O$, it follows that $\hat\kappa=\tilde\kappa^{-1}$, as stated.

\Halmos

\bigskip\bigskip\noindent%
The proof of Lemma \ref{Hauptsatz} is now almost complete.
For each $\O\in\K$, one has
$$K\dA(\O)K^*\subset\dA(\tilde\kappa(\O)),$$
and conversely,
$$\dA(\tilde\kappa(\O))=KK^*\dA(\tilde\kappa(\O))KK^*
\subset K\dA(\tilde\kappa^{-1}(\tilde\kappa(\O)))K^*=K\dA(\O)K^*,$$
so
$$K\dA(\O)K^*=\dA(\tilde\kappa(\O)),$$
and with $\xi:=\tilde\kappa(0)$ it follows from Lemma \ref{causal} that
$$K\dA(\O)K^*=\dA(\kappa\O+\xi)\qquad\forall\O\in\K.$$
That $\xi$ is unique, immediately follows from Lemma \ref{Landau2},
so the proof of Lemma \ref{Hauptsatz} is complete.\Halmos

\subsection*{Proof of Theorem \ref{mod sym} (i)}

It follows from Lemma \ref{Hauptsatz} that
there is a unique $\iota\in\Rd$ such that
$$J_{W_1}\dA(\O)J_{W_1}=\dA(j_1\O+\iota) \qquad\forall \O\in\K.$$
It remains to be shown that $\iota=0$.
Since $J$ is an involution, one has
$$x=j_1(j_1x+\iota)+\iota)=x+j_1\iota+\iota\qquad\forall x\in\Rd,$$
which gives $\iota=-j_1\iota$, hence $\iota_2=\dots=\iota_s=0$.
Furthermore, one has
$$\dA(W_1'+\iota)''=J_{W_1}\dA(W_1)''J_{W_1}=\dA(W_1)'$$
from Lemma \ref{Hauptsatz} and the Tomita-Takesaki theorem,
so on the one hand, it follows from Lemma \ref{Landau2} that
$$W_1'+\iota\subset W_1',$$
and on the other hand, locality implies
$$\dA(W_1')''\subset\dA(W_1)'=\dA(W_1'+\iota)''\subset\dA(W_1+\iota)',$$
so using Lemma \ref{Landau2} once more one finds
$$W_1'\subset W_1'+\iota,$$
arriving at $W_1'+\iota=W_1'$ and $\iota_0=\iota_1=0$, as stated.

\Halmos

\bigskip\bigskip\noindent%
In what follows, a well-known generalization of
Asgeirsson's Lemma will be used repeatedly. It is called the {\it double cone 
theorem} of Borchers and Vladimirov
\cite{Vla60,Bor61,Vla66,Bor97}. Below, it will be applied together
with the {\it edge of the wedge theorem} due to Bogoliubov
(cf., e.g., \cite{StW64,Vla66,Bor97}). For the
reader's convenience, both theorems are recalled here. For $\eps>0$,
$B_{\eps}$ will denote the open $\eps$-ball centered at the origin of
$\reals^2$, and $n$ will denote some natural number.

\begin{theorem}[Edge of the Wedge Theorem]\label{edge}
Let $C$ be a nonempty, open and convex
cone in $\reals^n$. For some $\eps>0$, assume
that $g_+$ is a function analytic in the tube $\reals^n+i(C\cap
B_\eps)$, and that $g_-$ is a function analytic in the tube
$\reals^n-i(C\cap B_\eps)$.
If there is an open region $\gamma\subset\reals^n$ where
$g_+$ and $g_-$ have a common boundary value in the sense of
distributions, then $g_+$ and $g_-$ are branches of a function $g$
which is analytic in
a complex neighbourhood $\Gamma$ of $\gamma$.

\end{theorem}

\begin{theorem}{\it
Given the assumptions and notation of Theorem \ref{edge}, let $c$ be any
smooth curve in $\gamma$ which has all its tangent
vectors in $C$. Then $g$ is analytic in a complex neighbourhood of the double cone
$(c+C)\cap(c-C)$.}

\end{theorem}
\noindent%
Another well known lemma that will be used repeatedly is the following
(cf. e.g., part (i) of Lemma 2.4.1 in \cite{Kuc99}).
\begin{lemma}\label{commutators}

Let $R\subset\Rd$ be a region that contains an open cone, and let
$A\in\aloc$ be a local observable such
that $\langle\gO,AB\gO\rangle=\langle\gO,BA\gO\rangle$
for all $B\in\dA(R)$. Then $A\in\dA(R)'$.

\end{lemma}

\subsection*{Proof of Theorem \ref{mod sym} (ii)}
In what follows, $e_0$ and $e_1$ denote the unit vectors pointing
into the $0$- and the $1$-direction, respectively.

For every $t\in\reals$, Theorem \ref{Hauptsatz} implies
the existence of a unique
$\xi(t)\in\Rd$ with
$$\modopt_{W_1}\dA(\O)\modopmt_{W_1}=\dA(\xi(t)+\Lambda_1(-2\pi t)\O)\qquad\forall\O\in\K.$$
By Corollary \ref{Landau2} it is clear that $\xi(t)+W_1=W_1$, so for
all $s\in\reals$, one has $\Lambda_1(-2\pi s)\xi(t)=\xi(t)$
and
\begin{align*}
\dA(\xi(s+t)+\Lambda_1(-2\pi&(t+s))\O)=
\Delta^{is}_{W_1}\modopt_{W_1}\dA(\O)\modopmt_{W_1}\Delta^{-is}_{W_1}\\
&=\dA(\xi(s)+\Lambda_1(-2\pi s)(\xi(t)+\Lambda_1(-2\pi t)\O))\\
&=\dA(\xi(s)+\Lambda_1(-2\pi s)\xi(t)+\Lambda_1(-2\pi(t+s))\O)\\
&=\dA(\xi(s)+\xi(t)+\Lambda_1(-2\pi(t+s))\O),
\end{align*}
so $\xi(s+t)=\xi(s)+\xi(t)$ follows from Lemma \ref{Landau2}. One now concludes
that $\xi(\lambda t)=\lambda\xi(t)$
for $\lambda\in{\bf Q}$, so $t\mapsto\xi(t)$ is ${\bf Q}$-linear.

Next we prove that the function $\reals\ni t\mapsto\xi(t)$ is continuous and, 
hence, $\reals$-linear.
As $\xi$ is additive, it is sufficient to prove continuity at $t=0$.
Assume $\xi$ were not continuous there, then there would exist a sequence
$(t_\nu)$, $\nu\in\naturals$, in $\reals$ that tends to zero, while $|\xi(t_\nu)|>\eps$ for some
$\eps>0$. Define the double cone
$$\O:=\left(-\mbox{$\frac{\eps}{3}$} e_0+V_+\right)\cap\left(\mbox{$\frac{\eps}{3}$}
e_0-V_+\right).$$ 
By the above results and locality, there is an
$N_\eps\in\naturals$ such that for any $A,B\in\dA(\O)$, one has
$$[\Delta_{W_1}^{it_\nu}A\Delta_{W_1}^{-it_\nu},B]=0\qquad\forall\nu>N_\eps.$$
But as $\Delta_{W_1}^{it}$ depends strongly continuously on $t$,
one concludes that $A$ and $B$ commute, and since $A$ and $B$ are
arbitrary elements of $\dA(\O)$, it follows that $\dA(\O)$ is
abelian. Additivity implies that $\aqloc''$ is abelian as
well, so $\H=\complex$ by irreducibility, which contradicts the
assumption that $\H$ is infinite-dimensional. It follows that
$\xi$ is continuous and, hence, $\reals$-linear, so there is a
$\xi\in\Rd$ with $\xi(t)=\xi t$ for all $t\in\reals$.

It remains to be shown that $\xi=0$. To this end, define the
double cone $\O:=(\rho e_1+V_+)\cap(\rho
e_1+\rho e_0-V_+)\subset W_1$ for some $\rho>0$. If one chooses
$\rho$ sufficiently small,
there are $a\in\Rd$ and $\eps,\delta>0$ such
that
\begin{quote}
(1) $\Lambda_1(-2\pi t)\O+t\xi-\delta t e_0\subset a+V_+$\, for
all $t\in[0,\eps]$;

(2) $\overline{\O}\not\subset a+V_+$.
\end{quote}
As an example, choose $a:=\rho e_1+\xi-|\xi|e_0$, where $|\xi|:=\sqrt{|\xi^2|}$.
Defining
$$f(t):=(\Lambda_1(-2\pi t)\rho e_1+t\xi-\delta te_0-a)^2,$$
one computes
$$f'(0)=2|\xi|(-2\pi\rho+|\xi|-\delta).$$
If one chooses
$\rho<\frac{|\xi|}{2\pi}$,
one can choose $\delta$ such that $0<\delta< -2\pi\rho+|\xi|$.
With this choice one has $f'(0)>0$, and as $f$ is smooth and satisfies
$f(0)=0$, there is an $\eps>0$ such that $f(t)\geq0$ for all $t\in[0,\eps]$,
which immediately implies Condition (1), whereas Condition (2)
follows from $f(0)=0$.

As the set $\bigcup_{0\leq t\leq\eps}(\Lambda_1(-2\pi t)\O+t\xi)$
is bounded, there is a $b\in\Rd$ such that
\begin{quote}
(3) $\Lambda_1(-2\pi t)\O+t\xi\subset b-V_+$\, for all
$t\in[0,\eps].$
\end{quote}
\begin{figure}[hth]\label{lorentzfirst}
\begin{center}
\input{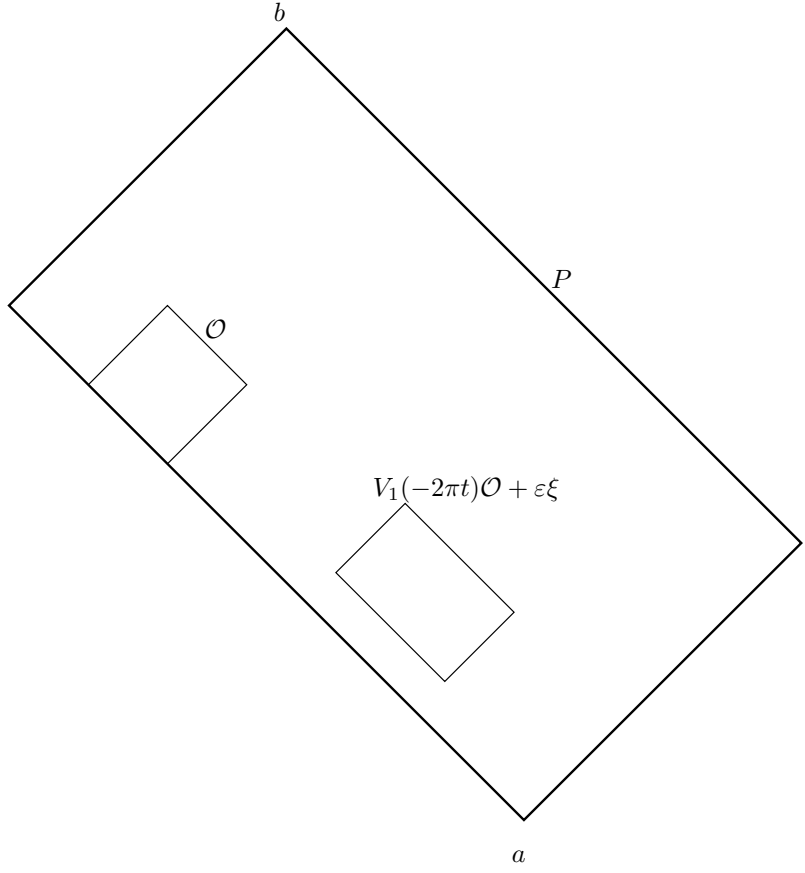}
\end{center}
{\bf \caption{The double cone $P$ in the proof of Thm. \ref{mod sym} (ii)}}
\end{figure}
Now denote $P:=(a+V_+)\cap(b-V_+)$ (Fig. 1), choose $A\in\dA(\O)$ and $B\in\dA(P')$,
denote by $e_0$ the unit vector in the time direction,
and consider the function $g_{A,B}$ defined by
$$\reals^2\ni(t,s)\mapsto g_{A,B}(t,s):=
\left<\gO,[B,U(se_0)\modopt_{W_1} A\modopmt_{W_1}
U(-se_0)]\gO\right>.$$ By Conditions (1) and (3), this function
vanishes in the closure of the open triangle $\gamma$ with
corners $(0,0)$, $(\eps,0)$ and $(\eps,-\delta\eps)$ (Fig. 2).
Clearly, $\gamma$ contains a smooth curve that joins $(0,0)$ to
$(\eps,-\delta\eps)$ and that has tangent vectors in the cone
$C:=\{(t,s)\in\reals^2:\,t\!>\!0,\, s\!<\!0\}$. It will be shown
that by the double cone theorem, $g_{A,B}$ vanishes in the whole
open rectangle $]0,\eps[\,\times\,]\!-\!\delta\eps,0[$.
\begin{figure}[hth]\label{zwei}
\begin{center}
\input{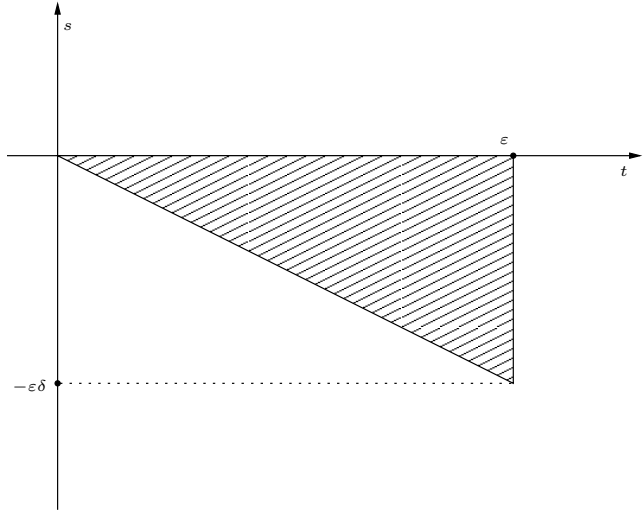}
\end{center}
{\bf \caption{Where $g_{A,B}$ vanishes in the proof of Thm. \ref{mod sym} (ii)}}
\end{figure}
Since $g_{A,B}$ is continuous, it follows that it even vanishes
in the closed rectangle \newline $[0,\eps]\,\times\,[-\delta\eps,0]$. Since
$B\in\dA(P')$ and $A\in\dA(\O)$ are arbitrary, Lemma \ref{commutators}
implies
that $\dA(\O-\delta\eps e_0)\subset\dA(P')'$.
But since by Condition (2), the double cone
$\O-\delta\eps e_0$ cannot be contained in $P$
no matter how small $\delta\eps$ is, this is
in conflict with Lemma \ref{Landau2}, so it follows that
$\xi=0$, which completes the proof.

It remains to be shown that the function $g_{A,B}$ fulfills the assumptions
of the double cone theorem. To this end, first note that
\begin{align*}
g_{A,B}&=\left<\gO,BU(se_0)\modopt A\gO\right>
-\left<\gO,A\modopmt U(-s e_0)B\gO\right>\\
&=\left<\gO,BU(se_0)\modopt A\gO\right>
-\overline{\left<\gO,B^*U(se_0)\modopt A^*\gO\right>}\\
&=:g_{+}(t,s)-g_{-}(t,s).
\end{align*}
Using elementary arguments from spectral theory it can be shown
that given any $\rho>0$, any
vector $\phi$ in the domain of $\Delta^{\rho}$ and any
$\psi\in\H$, the function $\reals\ni t\mapsto\langle\psi,\modopt\phi\rangle$
has an extension to a function
that is continuous on the strip $\{t\in\complex:\,
-\rho\leq\Im t\leq0\}$
and analytic on the interior of this strip  (cf. \cite{LiB92}, Lemma
8.1.10 (p. 351)).

As $\O\subset W_1$, the vectors $A\gO$ and $A^*\gO$ are in the domain of
$\Delta^{\frac{1}{2}}$, and it follows that for every $\psi\in\H$,
the functions $\reals\ni t\mapsto\langle\psi,\modopt A\gO\rangle$ and
$\reals\ni t\mapsto\overline{\langle\psi,\modopt A^*\gO\rangle}$ have
extensions that are continuous in the strips
$\{t\in\complex:\,-\frac{1}{2}\leq\Im t\leq0\}$ and
$\{t\in\complex:\,0\leq\Im\leq\frac{1}{2}\}$,
respectively, and that are analytic in the interior of these strips.

On the other hand, it follows from the spectrum condition that for
any two vectors $\phi,\psi\in\H$, the functions
$\reals\ni s\mapsto\langle\psi,U(s e_0)\phi\rangle$ and
$\reals\ni s\mapsto\overline{\langle\psi,U(s e_0)\phi\rangle}$ have
extensions that are continuous in the (complex)
closed upper and lower half plane,
respectively, and analytic in the interior of these half planes.

This proves that the function
$g_+$ has a continuous extension to the tube
${\bf T_+}:=\{(t,s)\in\complex^2:\,-1/2\leq\Im t\leq0,\,\Im s\geq0\}$
and that at every interior point of this strip, this extension
is analytic separately in $t$ and in
$s$. Using Hartogs' fundamental theorem stating that a function of
several complex variables is holomorphic if and only if it is
holomorphic separately in each of these variables \cite{Har06,Vla66},
it follows that $g_+$, as a function in two complex variables,
is analytic in the interior of ${\bf T_+}$.
It follows in the same way that $g_-$ has the
corresponding properties for the tube $-{\bf T_+}=:{\bf T_-}$.
The tubes ${\bf T_+}$ and ${\bf T_-}$ contain the smaller tubes
$\reals^2-i\overline{C\cap B_{\frac{1}{2}}}$
and $\reals^2+i\overline{C\cap B_{\frac{1}{2}}}$.

Since $g_+$ and $g_-$ coincide as continuous functions in the closure
of $\gamma$, they coincide as distributions in the open region
$\gamma$, and it follows from
the edge of the wedge theorem that they are branches of a function
$g$ that is analytic in a complex neighbourhood $\Gamma$ of $\gamma$.
But since $\gamma$ contains a smooth curve joining the points $(0,0)$ and
$(\eps,-\delta\eps)$
with tangent vectors in $C$, it follows from the double cone
theorem that the function $g$ is analytic in the region
$$((0,0)+C)\cap((\eps,-\delta\eps)-C)=]0,\eps[\,\times\,]-\delta\eps,0[.$$
This implies that $g_{A,B}$ vanishes in this region, which
is all that remained to be shown, so the proof is complete.\Halmos

\subsection*{Proof of Corollary \ref{mod sym+}}

If $J_+$ or $\modopt_+$ behave the way assumed in (i) or (ii),
respectively, the commutation relations recalled in the remark
preceding the corollary, together with Lemma \ref{Hauptsatz}, imply
that its geometrical action
can differ from the stated symmetry at most by a translation.
Since $V_+$ is Lorentz-invariant, $J_+$ and $\modopt_+$, $t\in\reals$,
commute with all $U(g),\,g\in L_+^{\uparrow}.$
However, there are no nontrivial translations that commute
with all $g\in L_+^{\uparrow}$.\Halmos

\subsection*{Proof of Lemma \ref{intermezzo}}

It follows from the Tomita-Takesaki Theorem that the modular
group under consideration leaves the algebras $\dA(W_1)''$ and
$\dA(W_1)'$ invariant. By wedge duality, it also leaves the
algebra $\dA(W_1')''=\dA(-W_1)''$ invariant. Borchers' commutation
relations now imply
$$\Delta_{W_1}^{i\eps}\dA(a\pm W_1)''\Delta_{W_1}^{-i\eps}
=\dA(\Lambda_1(-2\pi\eps)a\pm W_1)''.$$
$L(A)+W_1$ is a union of translates of $W_1$, so $(L(A)+W_1)^c$, being
an intersection of translates of $-\overline{W}_1$,
is a translate of $-\overline{W}_1$.
It follows that $\overline{(L(A)+W_1)^{cc}}$ is a translate of
$\overline{W}_1$. In particular,
$$\overline{(L(A)+W_1)^{cc}}=\bigcap\{a+\overline{W}_1:\,a\in\Rd,
\overline{(L(A)+W_1)^{cc}}\subset a+W_1\}.$$
But if $a\in\Rd$ is chosen such that $\overline{(L(A)+W_1)^{cc}}\subset a+W_1$,
Lemma \ref{Landau2} above and wedge duality imply
$A\in\dA(a+W_1')'=\dA(a+W_1)''$, so one finds
$$\bigcap\{a+\overline{W}_1:\,a\in\Rd,A\in\dA(a+W_1)''\}
\subset\overline{(L(A)+W_1)^{cc}},$$
and one concludes
\begin{align*}
  L(A_\eps)&\subset\bigcap\{a+
\overline{W}_1:\,a\in\Rd,\Delta_{W_1}^{i\eps}A\Delta_{W_1}^{-i\eps}\in\dA(a+W_1)''\}\\
  &=\bigcap\{a+
\overline{W}_1:\,a\in\Rd,A\in\Delta_{W_1}^{-i\eps}\dA(a+
W_1)''\Delta_{W_1}^{i\eps}\}\\
&=\bigcap\{a+
\overline{W}_1:\,a\in\Rd,A\in\dA(\Lambda_1(2\pi\eps)a+
W_1)''\}\\
&=\Lambda_1(-2\pi\eps)\bigcap\{a+
\overline{W}_1:\,a\in\Rd,A\in\dA(a+ W_1)''\}\\
&\subset\Lambda_1(-2\pi t)\overline{(L(A)+W_1)^{cc}}.
\end{align*}
The proof that $L(A_\eps)\subset\Lambda_1(-2\pi t)\overline{(L(A)-W_1)^{cc}}$
is completely analogous, so the proof of (i) is complete.

It remains to prove (ii) and (iii). We prove (iii); (ii) can be
established along precisely the same line of argument by
replacing $\modopt_{W_1}$ by $\modopmt_{W_1}$ and by exchanging,
respectively, $V_+$ and $-V_+$, $A$ and $A_\eps$ with one another.
Due to Borchers' commutation relations it suffices to consider
$A\in\dA(W_1)''$, which, as in the proof of Theorem \ref{mod sym} (ii),
will ensure that $A\gO\in D(\modop)$ in the following argument.

Assume that $L(A)\not\subset L(A_\eps)+\overline{V}_+$. Then one finds
an $a\in\Rd$ such that
\begin{quote}
(1)\quad $L(A_\eps)\subset a+V_+$, while

(2)\quad $L(A)\not\subset a+V_+$.
\end{quote}
This can be seen as follows. The assumption that
$L(A)\not\subset L(A_\eps)+\overline{V}_+$ and Statement (i) just proved imply that
there is a double cone $\O\subset L(A)$ such that $\overline{\O}$ and $L(A_\eps)$ are
spacelike separated, so there is a double cone $P\supset L(A_\eps)$ such that
$\O$ and $P$ are spacelike separated (cf., e.g., Prop. 3.8 (b) in \cite{TW97a});
choosing $a$ to be the lower tip of $P$, one arrives at both Conditions (1)
and Condition (2).

By Condition (1), $L(A_\eps)$ is a compact subset of the open set
$a+V_+$, and as $L(A_t)$ depends continuously on $t$ by assumption,
there exist $\sigma^\flat>0$ and $\delta>0$
such that
\begin{quote}
(1')\quad $L(A_{t})-\sigma^\flat e_0\subset a+V_+\qquad\forall
t\in[\eps-\delta,\eps]$,
\end{quote}
and this condition is, of course, equivalent to Condition (1).

Since $L(A_t)$ depends continuously on $t\in[0,\eps]$, the set
$\bigcup_{0\leq t\leq\eps}L(A_t)$ is bounded, so
one finds a $\sigma^\sharp\geq0$ such that
\begin{quote}
(3)\quad $L(A_t)+\sigma^\sharp e_0\subset a+V_+$ for all
$t\in[0,\eps]$,
\end{quote}
and for the same reason there is a $b\in\Rd$ such that
\begin{quote}
(4)\quad $L(A_t)+2\sigma^\sharp e_0\subset b-V_+$ for all
$t\in[0,\eps]$.
\end{quote}
Now define $P:=(a+V_+)\cap(b-V_+)$, and
for any $B\in\dA(P')$, consider -- as in the proof of
Proposition \ref{mod sym} -- the function $g_{A,B}$ defined by
$$\reals^2\ni(t,s)\mapsto g_{A,B}(t,s):=
\langle\gO,[B,U(se_0)A_t U(-se_0)]\gO\rangle.$$
Locality and Conditions (3) and (4) imply that this function
vanishes in the rectangle $[0,\eps]\times[\sigma^\sharp,2\sigma^\sharp]$,
and Condition (1') implies that it also vanishes in the
rectangle $[\eps-\delta,\eps]\times[-\sigma^\flat,\sigma^\sharp]$.
By the double cone theorem, $g_{A,B}$ vanishes
throughout the whole rectangle
$[0,\eps]\times[-\sigma^\flat,2\sigma^\sharp]$
(Fig. 3).
\begin{figure}[hth]\label{seconduniqueness}
\begin{center}
\input{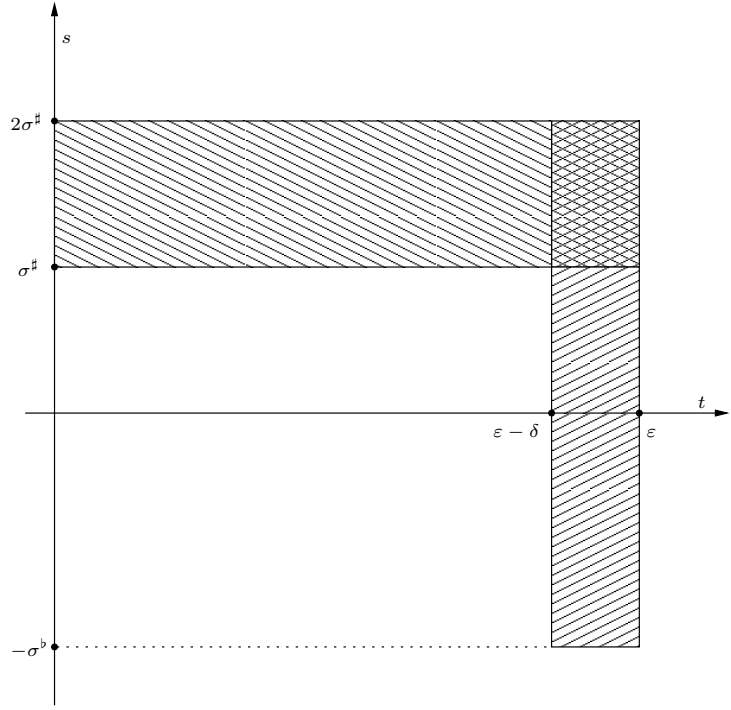}
\end{center}
{\bf \caption{where $g_{A,B}$ vanishes in the proof of Lemma \ref{intermezzo}}}
\end{figure}
In particular, one obtains $g_{A,B}(0,-\sigma^\flat)=0$ for all
$B\in\dA(P')$, so one can use Lemma
\ref{commutators} to conclude that
$A\in\dA(\sigma^\flat e_0+P')'$.
By the definition of $L(A)$, one finds
$$L(A)-\sigma^\flat e_0\subset\overline{P}\subset a+\overline{V}_+,$$
and as $\sigma^\flat>0$, this implies $L(A)\subset a+V_+$, which is 
in conflict with Condition (2) above and completes the proof.
\Halmos

\subsection*{Proof of Theorem \ref{second}}

Fix any $\rho>0$, and define the double cones
\begin{align*}
\O_1&:=(\rho(2 e_1+\,\,\,e_0)+V_+)\cap(\rho(2e_1+2e_0)-V_+),\\
\O_2&:=(\rho(2 e_1-2e_0)+V_+)\cap(\rho(2e_1+2e_0)-V_+),\\
\mbox{and}\qquad \O_3&:=(\rho(2e_1-3e_0)+V_+)\cap(\rho(2e_1+3e_0)-V_+),
\end{align*}
(Fig. 4)
\begin{figure}[hth]\label{lorsec}
\begin{center}
\input{lorsecklein.pstex_t}
\end{center}
{\bf \caption{The double cones $\O_1$, $\O_2$, and $\O_3$ in
the proof of Thm. \ref{second}}}
\end{figure}
and choose $A\in\dA(\O_1)$. As $L(A)\subset\overline{\O_1}$, it follows from
Lemma \ref{intermezzo} (i) and (ii) that
\begin{align*}
L(A_t)&\subset(\Lambda_1(-2\pi t)\rho\,(\mbox{$\frac{3}{2}$}e_1
+\mbox{$\frac{3}{2}$}e_0)+
\overline{W}_1)\\
&\qquad\cap(\Lambda_1(-2\pi
t)\rho\,(\mbox{$\frac{5}{2}$}e_1+\mbox{$\frac{3}{2}$}e_0)-\overline{W}_1)\\
&\qquad\cap(\rho(2e_1+2e_0)-\overline{V}_+)=:R_t,
\end{align*}
and there is an $\eps>0$ such that
$$R_t\subset\overline{\O_2}\qquad\forall t\in[0,\eps].$$
Note that by the linearity of the Lorentz boosts, $\eps$ does {\em not}
depend on $\rho$.
One now has $L(A_t)\subset\overline{\O}_2$ for all $A\in\dA(\O_1)$, and
with Corollary 5.4 in \cite{Kuc99}, it follows that
$$\modopt_{W_1}\dA(\O_1)\modopmt_{W_1}\subset\
\dA(\O_3')'\qquad\forall t\in[0,\eps].$$
Using Borchers' commutation relations, one finds
$$\modopt_{W_1}\dA(a+\O_1)\modopmt_{W_1}\subset\
\dA(\Lambda_1(-2\pi t)a+\O_3')'$$
for all $a\in\Rd$ and all $t\in[0,\eps]$.
Defining $x:=\rho(2e_1+e_0)$, 
$P_1:=\O_1-x$, and $P_3:=\O_3-x$, one obtains
$$\modopt_{W_1}\dA(a+P_1)\modopmt_{W_1}\subset\
\dA(\Lambda_1(-2\pi t)a+(x-\Lambda_1(-2\pi t)x)+P_3')'.$$
Note that the euclidean length of the vector
$x-\Lambda_1(-2\pi t)x$ is $\leq3\rho$
for all $t\in[0,\eps]$, as 
$\Lambda_1(-2\pi t)x\in R_t\subset\overline{\O}_2$ by the 
above choice of $\eps$.

Now choose any wedge $W\in\W$. As
$W\subset W+P_1$, it follows from wedge additivity that
$$\dA(W)''\subset\left(\bigcup_{a\in W}\dA(a+P_1)\right)''.$$
Define, for $\delta>0$, the wedges
$W^{(\delta)}:={\bf B}_\delta(W)''$,
where ${\bf B}_\delta(W)$ denotes the euclidean $\delta$-ball around
$W$, and $W^{(-\delta)}:=((W')^{(\delta)})'$, then it follows from isotony 
and wedge duality that
$$\left(\bigcup_{a\in W}\dA(a+P_3')'\right)''
\subset\dA(W^{(4\rho)})'',$$
and as the euclidean length of the vector $(\Lambda_1(-2\pi t)x-x)$ is 
$\leq 3\rho$, one arrives at
$$\left(\bigcup_{a\in W}\dA(a+(x-\Lambda_1(-2\pi t)x)+P_3')'\right)''
\subset\dA(W^{(7\rho)})''.$$
For $t\in[0,\eps]$, one now obtains
\begin{align*}
\modopt_{W_1}\dA(&\Lambda_1(2\pi t)W)''\modopmt_{W_1}\subset
\left(\bigcup_{a\in\Lambda_1(2\pi t)W}\modopt_{W_1}\dA(a+P_1)\modopmt_{W_1}\right)''\\
&\subset\left(\bigcup_{a\in\Lambda_1(2\pi t)W}\dA(
\Lambda_1(-2\pi t)a+(x-\Lambda_1(-2\pi t)x)+P_3')'\right)''\\
&\subset\dA(W^{(7\rho)})'',
\end{align*}
and as $W=(W^{(-7\rho)})^{(7\rho)}$,
this can be rewritten
$$\modopmt_{W_1}\dA(W)''\modopt_{W_1}\supset \dA(\Lambda_1(2\pi t)W^{(-7\rho)})'').$$
Using the fact that the transformations $\Lambda_1(2\pi t)$ are linear and, hence,
bounded maps in $\Rd$, which map the euclidean $7\rho$-ball onto some
bounded set with radius proportional to $\rho$, and 
using the facts that this radius continuously depends on $t\in[0,\eps]$,
that the interval $[0,\eps]$ is compact, and that 
$\eps$ does not depend on the choice of $\rho$,
one concludes that there is an $M>0$ which is independent from $\rho$ and 
satisfies
$$\Lambda_1(2\pi t)W^{(-7\rho)}\supset(\Lambda_1(2\pi t)W)^{(-M\rho)}
\qquad\forall t\in[0,\eps],$$
so with the above specifications of $\eps$ and $M$, one obtains
$$\modopmt_{W_1}\dA(W)''\modopt_{W_1}\supset \dA((\Lambda_1(2\pi t)W)^{(-M\rho)})''$$
for all wedges $W\in\W$ and all $\rho>0$. For each 
$A\in\aloc$, one now concludes
\begin{align*}
  L(A_t)&= \bigcap\{\overline{W}:\,W\in\W,\,\modopt_{W_1}A\modopmt_{W_1}
    \in\dA(W)''\} \\
    &=\bigcap\{\overline{W}:\,W\in\W,\,
    A\in\modopmt_{W_1}\dA(W)''\modopt_{W_1}\} \\
    &\subset\bigcap_{\rho>0}\bigcap\{\overline{W}:\,W\in\W,\,
    A\in\dA((\Lambda_1(2\pi t)W)^{(-M\rho)})''\} \\
    &=\bigcap_{\rho>0}\bigcap\{\overline{\Lambda_1(-2\pi t)X}:\,X\in\W,\,
    A\in\dA(X^{(-M\rho)})''\} \\
    &=\Lambda_1(-2\pi t)\bigcap_{\rho>0}\bigcap\{\overline{X}:
    \,X\in\W,\,A\in\dA(X^{(-M\rho)})''\} \\
    &=\Lambda_1(-2\pi t)\bigcap_{\rho>0}\bigcap\{\overline{X^{(M\rho)}}:
    \,X\in\W,\,A\in\dA(X)''\} \\
    &=\Lambda_1(-2\pi t)L(A).
\end{align*}
To prove the converse inclusion, one proves $L(A_t)\subset\Lambda_1(-2\pi t)$
for $t\in[-\eps,0]$ by mimicking the above argument: one defines the double cone
$$\O_1:=\rho(2 e_1-2e_0)+V_+)\cap(\rho(2e_1-e_0)-V_+),$$
keeps $\O_2$ and $\O_3$ as before, defines $x:=\rho(2e_1-e_0)$ and
proceeds like above with $t\in[-\eps,0]$, using Lemma \ref{intermezzo} (iii)
instead of Part (ii) of the same lemma. Now
having proved $L(A_t)\subset\Lambda_1(-2\pi t)L(A)$ for all $t\in[-\eps,\eps]$
and for all $A\in\aloc$, one concludes $L(A_t)=\Lambda_1(-2\pi t)L(A)$ for all
$t\in[-\eps,\eps]$ and for all $A\in\aloc$. 
As this immediately implies the statement for
all $t\in\reals$ and all $A\in\aloc$, the proof is complete.

\Halmos

\section{Conclusion}\label{conclusion}

By the above results, the modular group of a theory that does not
exhibit the Unruh effect acts in a completely ``non-geometric'' 
fashion, in the sense that it can neither preserve the net structure
nor act on the local observables in such a way that localization
regions evolve continuously. In particular, it cannot implement any
equilibrium dynamics in this case.

The above results imply that the only observer who can possibly
experience the vacuum in thermodynamical equilibrium is the uniformly
accelerated one (whose acceleration may, of course, be zero).
Physically, this result reflects the fact that any non-uniformly
accelerated observer would feel nonstationary inertial forces destroying
any thermodynamical equilibrium, while the constant acceleration felt by
a uniformly accelerated observer does not affect thermodynamical equilibrium
provided the theory exhibits the Unruh effect.

The first results similar to the above ones have been obtained by Araki and
by Keyl \cite{Ara92,Key93}.
These authors avoid the spectrum condition and assume stronger
a priori restrictions on the possible geometric behaviour instead.
Recently, more results in this spirit have been found by
Buchholz et al. and by Trebels \cite{BDFS98,Dre96,Flo99,Tre97}. 
One aim of these approaches is
to obtain new insight on quantum fields on curved spacetimes by
avoiding the spectrum condition. So far, results have been obtained
for de Sitter, Anti-de Sitter, and certain Robertson-Walker
spacetimes \cite{BDFS98,BFS99,BMS00}.

For the vacuum states in Minkowski space considered above, the
spectrum condition is a reasonable physical assumption. The
assumptions made above on the possible geometric behaviour of the
modular objects (in particular those made in the first uniqueness
theorem) are less restrictive than those made in any of the other
approaches, since a small class of regions, namely, the double cones,
is assumed to be mapped into an extremely large class of regions,
namely, the open sets. In this sense the above results are, at present,
the most general uniqueness results in Minkowski space that point
towards the Unruh effect and modular \pct-symmetry.

Even more than a uniqueness result
can be found if conformal symmetry holds in addition to our
above Conditions (A) through (C). In this case,
the whole representation of the conformal group
arises from the modular objects of the theory, and
in particular, the Bisognano-Wichmann symmetries can be established
\cite{BGL93}.

\section*{Appendix. A Remark on the Continuity of $t\mapsto L(A_t)$}

In the discussion of the second uniqueness theorem it was assumed that
$L(A_t)$ depends continuously on $t$ for $t\in[0,\eps]$ 
in the sense that for each sequence 
$(t_\nu)_{\nu\in\naturals}$ tending to a $t_\infty\in[0,\eps]$,
the localization region $L(A_{t_{\infty}})$ consists precisely of all 
accumulation points of sequences $(x_\nu)_{\nu\in\naturals}$
with $x_\nu\in L(A_{t_\nu})$. In this appendix we show that this notion
of convergence, which we refer to as {\em pointwise convergence}, 
is equivalent to the convergence according to a metric
first considered by Hausdorff, which 
one can introduce on the set $\C$ of compact convex subsets of $\Rd$
by defining, for any two such sets $K,L\in\C$,
$$\delta_H(K,L):=\inf\{\delta>0:\,K\subset{\bf B}_\delta(L)\,\mbox{and}\,
L\subset{\bf B}_\delta(K)\}$$
(cf. Problem 4D (p. 131) in \cite{Kel}).
It is evident that continuity of $[0,\eps]\ni t\mapsto L(A_t)$ 
with respect to this metric, which we refer to as {\em
uniform continuity}, implies the
pointwise continuity for this map. Conversely, 
one can also show that pointwise continuity implies uniform
continuity for $t\mapsto L(A_t)$.

To prove this indirectly, assume that
$t\mapsto L(A_t)$ is pointwise continuous for
$t\in[0,\eps]$ and that this map is not continuous with respect to 
Hausdorff's metric. Then there exists a $\rho>0$ and  
a sequence $(t_\nu)_{\nu\in\naturals}$
of points in $[0,\eps]$ which converges to a point $t_\infty\in[0,\eps]$
and has the property that 
$$\delta_H(L(A_{t_\nu}),L(A_{t_\infty}))\geq\rho.$$
On the other hand, 
there is a subsequence $(s_\nu)_{\nu\in\naturals}$ of 
$(t_\nu)_{\nu\in\naturals}$ with the property that all
$L(A_{s_\nu})$ have nonempty intersection with 
${\bf B}_\rho(L(A_{t_\infty}))$, as otherwise 
$L(A_{t_\infty})$ would be empty by the assumption of 
pointwise continuity.

As $\delta_H(L(A_{s_\nu}),L(A_{t_\infty}))\geq\rho$,
there exists a sequence $(x_\nu)_{\nu\in\naturals}$ such that
the euclidean distance $\delta(x_\nu,L(A_{t_\infty}))$ between 
$x_\nu$ and $L(A_{t_\infty})$ is $\geq\rho/2$
for all $\nu\in\naturals$,
and as all $L(A_{s_\nu})$ are convex sets with a nonempty intersection with 
${\bf B}_\rho(L(A_{t_\infty}))$, this sequence can be chosen such that it is
bounded and, hence, has an accumulation point $\tilde x$.
As $\delta(x_\nu,L(A_{t_\infty}))\geq\rho/2$ for all $\nu\in\naturals$,
one finds $\delta(\tilde x,L(A_{t_\infty})\geq\rho/2$, so
$\tilde x\notin L(A_{t_\infty})$.
But this contradicts the assumption that $t\mapsto L(A_t)$ is pointwise
continuous and proves that this map is pointwise continuous if and only if
it is uniformly continuous, as stated.

It is now easy to see that $\bigcup_{t\in[0,\eps]}L(A_t)$ is
bounded, as stated in the text. Namely,
the function $[0,\eps]\ni t\mapsto
\delta_H(L(A),L(A_t))$ is continuous and, hence, has a maximum
$\rho>0$ in the compact interval $[0,\eps]$. It follows that 
$\bigcup_{t\in[0,\eps]}L(A_t)
\subset{\bf B}_\rho(L(A))$, which is a bounded set. 

\bigskip
\subsection*{Acknowledgements} It was an important help that D. Arlt and 
N. P. Landsman read the manuscript carefully. 

This research was funded by the Deutsche Forschungsgemeinschaft, a
Feodor-Lynen grant of
the Alexander von Humboldt foundation, and a 
Hendrik Casimir-Karl Ziegler award
of the Nordrhein-Westf\"alische Akademie der Wissenschaften. 

The idea to reinitiate the project originated during a stay in 1997 at the 
Erwin-Schr\"odinger
Institute for Mathematical Physics at Vienna. Helpful discussions there 
with S. Trebels and D. Guido are gratefully acknowledged.


\begin{thebibliography}{******}

\bibitem{Ale50} Alexandrov, A. D.: On Lorentz transformations. Uspekhi
  Mat. Nauk. {\bf 5}, No. 3 (37), 187 (1950)

\bibitem{Ale75} Alexandrov, A. D.: Mappings of Spaces with Families of
  Cones and Space-Time Transformations. Annali di matematica {\bf
  103}, 229-257 (1975)

\bibitem{AlO53} Alexandrov, A. D., Ovchinnikova, V. V.: Notes on the
  foundations of relativity theory. Vestnik Leningrad Univ. {\bf 14},
  95 (1953)

\bibitem{Ara92} Araki, H.: Symmetries in a Theory of Local Observables
and the Choice of the Net of Local Algebras, Rev. Math. Phys.,
Special Issue, 1-14 (1992)

\bibitem{Ara99} Araki, H.: {\it Mathematical Theory of Quantum Fields}.
Oxford: Oxford University Press, 1999

\bibitem{BaW92} Baumg\"artel, H., Wollenberg, M.: {\it Causal Nets of Operator
Algebras}. Berlin: Akademie-Verlag, 1992

\bibitem{BW75} Bisognano, J. J., Wichmann, E. H.: On the Duality
 Condition for a Hermitian Scalar Field. J. Math. Phys. {\bf 16},
 985-1007 (1975)

\bibitem{BW76} Bisognano, J. J., Wichmann, E. H.: On the Duality
Condition for Quantum Fields. J. Math. Phys. {\bf 17}, 303 (1976)

\bibitem{Bor61} Borchers, H.-J.: \"Uber die Vollst\"andigkeit
  lorentzinvarianter Felder in einer zeitartigen R\"ohre. Nuovo
  Cimento {\bf 19}, 787-796 (1961)

\bibitem{Bor65} Borchers, H.-J.: On the Vacuum State in Quantum Field Theory,
II. Commun. Math. Phys. {\bf 1}, 57 (1965)


\bibitem{Bor92} Borchers, H.-J.: The CPT-Theorem in Two-Dimensional
Theories of Local Observables. Commun. Math. Phys. {\bf 143}, 315-332
(1992)

\bibitem{Bor97} Borchers, H.-J.: Translation Group and Particle
  Representations in Quantum Field Theory. Berline, Heidelberg: Springer, 
  1996

\bibitem{Bor97a} Borchers, H.-J.: On Poincar\'e transformations and
  the modular group of the algebra associated with a wedge.
{ Lett. Math. Phys.} {\bf 46}, 295-301 (1998)

\bibitem{Bor99} Borchers, H.-J.:
On the Revolutionization of Quantum Field Theory
by Tomita's Modular Theory. { J. Math. Phys.} {\bf 41},
3604-3673 (2000)

\bibitem{BoH72} Borchers, H.-J., Hegerfeldt, G. C.: The Structure of
  Space-Time Transformations. Commun. Math. Phys. {\bf 28},
259-266 (1972)

\bibitem{BGL93} Brunetti, R., Guido, D., Longo, R.: Modular Structure
 and Duality in Conformal Quantum Field Theory. Commun. Math. Phys.
{\bf 156}, 201-219 (1993)


\bibitem{Buc75} Buchholz, D.: Collision Theory for Massless Fermions.
Commun. Math. Phys. {\bf 42}, 269-279 (1975)

\bibitem{Bu75a} Buchholz, D.: Collision Theory for Waves in Two
Dimensions and a Characterization of Models with Trivial S-Matrix.
Commun. Math. Phys. {\bf 45}, 1-8 (1975)

\bibitem{Buc77} Buchholz, D.: Collision Theory of Massless Bosons.
Commun. Math. Phys. {\bf 52}, 147-173 (1977)

\bibitem{Bu77a} Buchholz, D.: On the Structure of Local Quantum Fields
with Non-Trivial Interaction. In: Proceedings
of the International Conference on Operator Algebras, Ideals and
Their Applications in Theoretical Physics, Leipzig, 1977. Stuttgart:
Teubner, 1978

\bibitem{BDFS98} Buchholz, D., Dreyer, O., Florig, M., Summers, S. J.:
  Geometric Modular Action and spacetime Symmetry Groups. { Rev. Math. Phys.}
{\bf 12}, 475-560 (2000)

\bibitem{BFS99} Buchholz, D. Florig, M., Summers, S. J.: Hawking-Unruh
Temperature and Einstein Causality in Anti-de Sitter Space-Time.
{ Class. Quant. Grav.} {\bf 17}, L31-L37 (2000)

\bibitem{BF77} Buchholz, D., Fredenhagen, K.: Dilations and
  interaction. { J. Math. Phys.} {\bf 18}, 1107-1111 (1977)

\bibitem{BMS00} Buchholz, D., Mund, J., Summers, S. J.: Transplantation
of Local Nets and Geometric Modular Action on Robertson-Walker 
Space-Times. Preprint, {\tt hep-th/0011237}

\bibitem{BuS93} Buchholz, D., Summers, S. J.: An Algebraic Characterization of
Vacuum States in Minkowski Space. Commun. Math. Phys. {\bf 155}, 449-458
(1993)

\bibitem{Dav95} Davidson, D. R.: Modular Covariance and the Algebraic
PCT/Spin-Statistics Theorem. Preprint, {\tt hep-th/9511216}

\bibitem{Dre96} Dreyer, O.: Das Prinzip der geometrischen modularen Wirkung im
de Sitter-Raum. diploma thesis, University of Hamburg, 1996

\bibitem{Flo98} Florig, M.: On Borchers' Theorem. { Lett Math. Phys.}
{\bf 46}, 289-293 (1998)

\bibitem{Flo99} Florig, M.: Geometric Modular Action. PhD-thesis, University
of Florida, Gainesville, 1999

\bibitem{GL94} Guido, D., Longo, R.: An Algebraic Spin and Statistics
Theorem. Commun. Math. Phys. {\bf 172}, 517-534 (1995)

\bibitem{GL95} Guido, D., Longo, R.: The Conformal Spin and Statistics
Theorem. Commun. Math. Phys. {\bf 181}, 11-36 (1996)

\bibitem{Haa92} Haag, R.: {\it Local Quantum Physics}. Berlin: Springer, 1992

\bibitem{Har06} Hartogs, F.: Zur Theorie der Funktionen mehrerer
  komplexer Ver\"anderlicher, insbesondere \"uber die Darstellung
  derselben durch Reihen, welche nach Potenzen einer Ver\"anderlichen
  fortschreiten. { Math. Ann.} {\bf 62}, 1-88 (1906)

\bibitem{Kel} Kelley, J. L.: {\it General Topology}. New York: 
van Nostrand, 1955

\bibitem{Key93} Keyl, M.: Remarks on the relation between causality
and quantum fields. { Class. Quantum Grav.} {\bf 10}, 2353-2362
(1993)

\bibitem{Kuc95} Kuckert, B.: A New Approach to Spin \& Statistics.
{ Lett. Math. Phys.} {\bf 35}, 319-335 (1995)

\bibitem{Kuc97} Kuckert, B.: Borchers' Commutation Relations and
  Modular Symmetries in Quantum Field Theory. { Lett. Math. Phys.}
  {\bf 41}, 307-320 (1997)

\bibitem{Kuc98} Kuckert, B.: Spin \& Statistics, Localization Regions,
and Modular Symmetries in Quantum Field Theory. PhD-thesis, Hamburg
1998, DESY-thesis 1998-026

\bibitem{Kuc99} Kuckert, B.: Localization Regions of Local Observables.
Commun. Math. Phys. {\bf 215}, 197-216 (2000)

\bibitem{LiB92} Li Bing-Ren: {\it Introduction to Operator Algebras}. Singapore:
  World Scientific, 1992

\bibitem{Lon96} Longo, R.: On the spin-statistics relation for
  topological charges. In: Doplicher, S., Longo, R., Roberts, J. E.,
  Zsido, L. (eds.): {\it Operator Algebras and Quantum Field Theory}.
  Proceedings of the conference at the Accedemia Nazionale dei Lincei,
  Rome 1996. Cambridge, MA: International Press, 1997

\bibitem{MS69} Mack, G., Salam, A.: Finite-Component Field
  Representations of the Conformal Group. { Ann. Phys.} {\bf 53},
  174-202 (1969)


\bibitem{Mun98} Mund, J.: Quantum Field Theory of Particles with
Braid Group Statistics in 2+1 dimensions. PhD-thesis, Freie
Universit\"at Berlin, 1998

\bibitem{RS61} Reeh, H., Schlieder, S.: Bemerkungen zur Unit\"ar\"aquivalenz
von lorentzinvarianten Feldern. { Nuovo Cimento} {\bf 22}, 1051 (1961)

\bibitem{StW64} Streater, R. F., Wightman, A. S.: {\it PCT, Spin \& Statistics,
and All That}, New York: Benjamin, 1964

\bibitem{Tak70} Takesaki, M.: Tomita's Theory of Modular Hilbert
Algebras and Its Applications. Lecture Notes in Mathematics
{\bf 128}, New York: Springer, 1970

\bibitem{TW97a} Thomas, L. J., Wichmann, E. H.: Standard forms of
  local nets in quantum field theory. { J. Math. Phys.} {\bf 39},
  2643-2681 (1998)

\bibitem{Tre97} Trebels, S.: PhD-thesis. G\"ottingen 1997,
cf. also \cite{Bor99}

\bibitem{Unr76} Unruh, W. G.: Notes on black hole evaporation. Phys. Rev.
{\bf D14}, 870-892 (1976)

\bibitem{Vla60} Vladimirov, V. S.: The construction of envelopes of
  holomorphy for domains of a special type (in Russian).
  { Doklady Akad. Nauk SSSR} {\bf 134}, 251-254 (1960)

\bibitem{Vla66} Vladimirov, V. S.: Methods of the Theory of Functions
  of Many Complex Variables. Cambridge, MA: M. I. T. Press, 1966

 \bibitem{Wie92} Wiesbrock, H.-W.: A Comment on a Recent Work of Borchers.
Lett. Math. Phys. {\bf 25}, 157-159 (1992)

\bibitem{Yng94} Yngvason, J.: A Note on Essential Duality.
 { Lett. Math. Phys.} {\bf 31}, 127-141 (1994)

\bibitem{Zee64} Zeeman, E. C.: Causality Implies the Lorentz Group.
{ J. Math. Phys.} {\bf 5}, 490-493 (1964)
\end{thebibliography}
\end{document}